\documentclass[12pt]{article}
\usepackage{amsfonts,cite}

\newfam\scrfam
\batchmode\font\twelvescr=rsfs12 \errorstopmode
\ifx\tenscr\nullfont
        \newcal{\cal}
\else	\font\sevenscr=rsfs7 
	\font\fivescr=rsfs5 
	\skewchar\twelvescr='177 \skewchar\sevenscr='177 \skewchar\fivescr='177
	\textfont\scrfam=\twelvescr \scriptfont\scrfam=\sevenscr
	\scriptscriptfont\scrfam=\fivescr
	\def\scr{\fam\scrfam}
	\def\newcal{\scr}
\fi

\newcommand{\be}{\begin{equation}}
\newcommand{\ee}{\end{equation}}
\newcommand{\bea}{\begin{eqnarray}}
\newcommand{\eea}{\end{eqnarray}}
\newcommand{\ba}{\begin{array}}
\newcommand{\ea}{\end{array}}

\newcommand{\al}{\alpha}
\newcommand{\ga}{\gamma}
\newcommand{\Ga}{\Gamma}

\newcommand{\ka}{\kappa}
\newcommand{\de}{\delta}
\newcommand{\ep}{\epsilon}

\newcommand{\la}{\lambda}

\newcommand{\Om}{\Omega}
\newcommand{\ze}{\zeta}

\newcommand{\La}{\Lambda}

\newcommand{\Ups}{\Upsilon}
\newcommand{\vth}{\vartheta}
\newcommand{\Z}{\mathbb{Z}}
\newcommand{\R}{\mathbb{R}}

\newcommand{\D}{{\rm d}}
\newcommand{\DD}{{\rm D}}

\newcommand{\id}{\hbox{1\kern-.27em l}}
\newcommand{\sid}{\hbox{\scriptsize1\kern-.27em l}}

\newcommand{\pa}{\partial}
\newcommand{\rar}{\rightarrow}
\newcommand{\non}{\nonumber}
\newcommand{\we}{\kern-.1em\wedge\kern-.1em}
\newcommand{\scal}{\kern-.13em\cdot\kern-.13em}
\newcommand{\im}{{\rm Im}}
\newcommand{\re}{{\rm Re}}
\newcommand{\cC}{{\newcal C}}
\newcommand{\cH}{{\newcal H}}
\newcommand{\cF}{{\newcal F}}
\newcommand{\cO}{{\cal O}}

\newcommand{\cU}{{\newcal U}}
\newcommand{\cK}{{\newcal K}}
\newcommand{\bcK}{\bar{\newcal K}}

\newcommand{\bcH}{\bar{\newcal H}}
\newcommand{\bcF}{\bar{\newcal F}}

\newcommand{\bcU}{\bar{\newcal U}}

\newcommand{\bE}{{\bar{E}}}
\newcommand{\half}{\mbox{$\frac{1}{2}$}}
\newcommand{\third}{\mbox{$\frac{1}{3}$}}
\newcommand{\fourth}{\mbox{$\frac{1}{4}$}}
\newcommand{\sixth}{\mbox{$\frac{1}{6}$}}
\newcommand{\II}{I\kern-.09em I}

\begin{document}

\vspace*{-6mm}

\rightline{\vbox{\footnotesize
\hbox{DAMTP-1999-57}
\hbox{\tt hep-th/9905019}
}}

\vspace{3mm}

\begin{center}
{\Large\sf Towards a manifestly SL(2,$\Z$)-covariant action\\ 
for the type {\II}B {\Large $(p,q)$} super-five-branes }

\vskip 5mm

Anders Westerberg\footnote{\tt A.Westerberg@damtp.cam.ac.uk}
and Niclas Wyllard\footnote{\tt N.Wyllard@damtp.cam.ac.uk} \vspace{5mm}\\
{\em DAMTP, University of Cambridge,\\ Silver Street, Cambridge CB3 9EW, UK}

\end{center}
 
\vskip 5mm
 
\begin{abstract}
\noindent We determine a manifestly SL(2,$\Z$)-covariant $\ka$-symmetric 
action for the type {\II}B $(p,q)$ five-branes 
as a perturbative expansion in the world-volume field strengths 
within the framework
where the brane tension is generated by a world-volume field.
In this formulation the Lagrangian is expected to be polynomial;
we construct the $\ka$-invariant action to fourth order in the
world-volume field strengths. 
\end{abstract}

\setcounter{equation}{0}
\section{Introduction}

Type {\II}B superstring theory is known to have a non-perturbative
SL(2,$\Z$) symmetry~\cite{Hull:1995} under which the 
$p$-branes of the
theory fall into representations. The strings transform in a doublet 
(``$(p,q)$ strings'') \cite{Schwarz:1995,Witten:1995b}, whereas the 
three-brane is an SL(2,$\Z$) singlet~\cite{Duff:1991}. The
five-branes again belong to a doublet as argued in ref.~\cite{Witten:1995b}. 
Supergravity solutions for these $(p,q)$ five-branes 
were constructed in ref.~\cite{Lu:1998}. In addition, there are
seven-branes in the theory 
which should transform in a triplet under SL(2,$\Z$) \cite{Meessen:1998}. 
There exist formulations of type {\II}B supergravity (to which the
above branes couple) in which the SL(2) symmetry is manifest
\cite{Howe:1984,Dall'Agata:1998}, a fact that can be exploited to construct 
world-volume actions for the $p$-branes of the type {\II}B theory displaying
manifest SL(2) symmetry. This programme has been completed for the
strings in refs~\cite{Townsend:1997,Cederwall:1997} and for the three-brane in
ref.~\cite{Cederwall:1998a}. So far, however, such formulations are lacking
for the higher-dimensional branes. It is the purpose of this note to 
investigate the case of a manifestly SL(2)-covariant action for the
five-brane doublet. 

The world-volume theory of a $(p,q)$ five-brane should 
be described on shell by a six-dimensional vector super-multiplet.
However, in order to make the SL(2) symmetry manifest, one  
needs to introduce additional dynamical fields in the action. 
Our treatment will be within the framework where the brane tension is 
generated by a world-volume field---in the present 
case a complex six-form field strength. In this formulation the
Lagrangian is expected to be a polynomial function of the gauge-invariant
world-volume field strengths. 
To regain the correct counting for the degrees of freedom, auxiliary duality 
relations are then imposed at the level of the equations of motion.
For the three-brane, for instance, a complex world-volume two-form 
field strength (or two real ones) satisfying a non-linear
duality relation is required \cite{Cederwall:1998a}. 

In ref.~\cite{Cederwall:1998a}, some aspects of a manifestly 
covariant formulation of a $(p,q)$ five-brane action were presented; 
we continue this study here using the constructive method of
refs~\cite{Cederwall:1998a,Cederwall:1998b,Westerberg:1999}. The
analysis is complicated by the fact that the tension form is
complex and by a ``non-canonical'' structure of the auxiliary duality
relations, forcing us to resort to a perturbative treatment.
The main result of our investigations is the determination of the
$\ka$-symmetric, manifestly SL(2)-covariant action and the
associated projection operator for the type {\II}B $(p,q)$ five-branes
to fourth order in the world-volume field strengths. 

In the next section we discuss some facts about the background type {\II}B 
supergravity theory and some general features of the world-volume theory of 
the five-brane doublet. In section~\ref{results} we then discuss the method 
used to construct the action and present our results. 
Finally, we list our conventions in a short appendix.

\setcounter{equation}{0}
\section{Preliminaries}
\label{prel}

The type {\II}B supergravity theory in ten dimensions 
\cite{Schwarz:1983,Howe:1984} is chiral and has a U(1) R-symmetry.
In the complex superspace formulation~\cite{Howe:1984} the two Majorana--Weyl 
spinorial superspace coordinates are combined into a complex Weyl spinor. 
The theory, furthermore, has an SL(2,$\R$) symmetry at the classical level, 
which is broken down to SL(2,$\Z$) by non-perturbative quantum effects. 
By gauging the U(1) R-symmetry it is possible to formulate the theory in a 
way which makes the SL(2) symmetry manifest. In this formulation the scalars 
of the theory belong to the coset space SL(2,$\R$)/U(1).
More precisely, the scalars form a $2{\times}2$ matrix 
\be
\label{scalmatr}
\left(\ba{cc} \cU^1 & \bcU^1 \\ \cU^2 & \bcU^2 \ea \right)
\ee 
on which SL(2,$\R$) acts from the left and U(1) acts locally
from the right, both group actions leaving  invariant the constraint
$\frac{i}{2}\ep_{rs}\cU^r\bcU^s = 1$ (here $\ep_{12}=-1$). 
From the components of the above matrix one can construct 
the one-forms
\be
Q =\half\,\ep_{rs}\,\D\cU^r\bcU^s\,, \qquad
P = \half\,\ep_{rs}\D\cU^r\cU^s\,,
\ee
which have special significance~\cite{Howe:1984}. Both are 
SL(2,$\R$)-invariant, whereas under 
the local U(1) transformation $\cU^r\rar\cU^re^{i\vth}$ they transform as 
$Q \rar Q+\D\vth$ and $P\rar P e^{i2\vth}$, respectively. 
Hence, the real one-form $Q$ is a U(1) connection, while $P$ has U(1) 
charge $+2$ (the U(1) charge of $\cU^r$ is normalised to +1). 
They furthermore satisfy 
\be
\label{mceq}
\D Q-i\,P\we\bar{P} = 0\,, \qquad \D P-2i\,P\we Q = 0\,,
\ee
the second equation showing that $P$ is U(1)-covariantly constant 
(the U(1)-covariant derivative is $\DD=\D-i e\,Q$, where $e$ denotes the
U(1) charge, and acts from the 
right). 
The two physical scalars of the theory are encoded in the projective invariant
$\cU^1/\cU^2$, which in our conventions later will be identified with
$-\bar{\tau}=-C_0+i\,e^{-\phi}$.

In addition to the vielbein and the scalars, there are a four-form potential whose field strength is non-linearly 
self-dual and two two-form potentials. When dealing with
$p$-branes with $p\!>\!3$ one needs to use a formulation of the supergravity 
theory in which the Poincar\'e duals to all field strengths are included on 
the same footing as the original forms. In the present case we
therefore also have 
two six-form potentials (for the seven-branes the eight-form potentials 
become important too). 
The SL(2) doublet $\cU^r$ discussed above serves
 as a bridge between quantities 
transforming in the fundamental representation of SL(2) and
SL(2)-invariant (complex) 
quantities which are charged under the gauged U(1) R-symmetry. 
Examples include the two- and six-form potentials above,
which can be expressed in terms of the SL(2) invariant 
forms $\cC_2=\cU^rC_{2;r}$ and
$\cC_6 = \cU^rC_{6;r}$, both of U(1) charge $+1$ (here $C_{2;1}=B_2$,
$C_{2;2}=C_2$ and similarly for $C_{6;r}$). We use calligraphic
letters to denote complex quantities with U(1) charge $+1$. Complex
conjugation is indicated with a bar and the corresponding quantities
have U(1) charge $-1$. The background field strengths which we need are
\bea
\cH_3 &=& \cU^r\,\D C_{2;r}\,, \non \\
H_5 &=& \D C_4 + \half\, \im (\cC_2\,\bcH_3)\,, \non \\
\cH_7 &=& \cU^r\,\D C_{6;r} + x\,\cC_2\,H_5 - (1{-}x)\,C_4\,\cH_3 +
\half(\third{-}x)\,\im(\cC_2\,\bcH_3)\,\cC_2\,,
\eea
where, following ref.~\cite{Westerberg:1999}, we have introduced a free 
parameter $x$ in the definition of $\cH_7$.
These fields satisfy the Bianchi identities 
\bea
\DD \cH_3 + i\, \bcH_3\,P &=& 0\,, \non \\
\D H_5 - \mbox{$\frac{i}{2}$}\, \cH_3 \, \bcH_3 &=& 0\,, \non \\
\DD \cH_7  + i\, \bcH_7\,P + \cH_3\, H_5 &=& 0\,. 
\eea
The constraints which have to be imposed in the superspace approach
are at dimension~0 
\bea
\label{dim0}
{T_{\al\bar{\beta}}}^a = {T_{\bar{\al}\beta}}^a &=&
i\,(\ga^a)_{\al\beta}\,, \non \\
\cH_{a\al\beta} &=& 2\,(\ga_a)_{\al\beta}\,, \non \\
H_{abc\bar{\al}\beta} = -H_{abc\al\bar{\beta}} &=&
(\ga_{abc})_{\al\beta}\,, \non \\
\cH_{abcde\al\beta} &=& 2i\,(\ga_{abcde})_{\al\beta}\,.
\eea
Here the barred indices on the left-hand sides refer to components 
corresponding to the basis form $E^{\bar{\al}}=\overline{E^{\al}}$;
since barred and un-barred indices are of the same type, the bars have been 
dropped on the right-hand side (see also the appendix).

For fermionic backgrounds one also needs the dimension 1/2 constraints
\bea 
\label{dim1/2}
P_\al &=& 2i\,\La_\al\,, \non \\
\cH_{ab\bar{\al}} &=& 2i\,(\ga_{ab}\,\La)_{\al}\,, \non \\
\cH_{\hbox{\scriptsize\it abcdef}\bar{\al}} &=& 2\,
(\ga_{\hbox{\scriptsize\it abcdef}}\,\La)_{\al}\,,
\eea
where $\La_\al$ is the dilatino superfield of U(1) charge $+\frac{3}{2}$.
The two sets of constraints given above put the theory on-shell. 
Note that we have only displayed the non-vanishing components that are
relevant for our calculations. An expedient way to obtain the constraints 
for $\cH_7$ is by translating the results of ref.~\cite{Cederwall:1996c} 
into the complex formulation used here (taking into account the sign misprint 
corrected in ref.~\cite{Westerberg:1999}). The particular choice of
gauge we have made use of in converting between the real and complex 
formulations is $\cU^1=-e^{\frac{1}{2}\phi}C_0+ie^{-\frac{1}{2}\phi}$
and $\cU^2=e^{\frac{1}{2}\phi}$.

Let us next consider the world-volume gauge field content of the $(p,q)$ 
five-brane theory. In order to be able to 
formulate the world-volume action in a 
manifestly SL(2)-covariant manner one needs the following gauge-invariant
world-volume forms:
\bea
\cF_2 &=& \cU^r \D A_{1;r} - \cC_2\,, \non \\
F_4 &=& \D A_3 - C_4 + \half\,\im(C_2\, \bcF_2)\,, \non \\
\cF_6 &=& \cU^r \D A_{5;r} - \cC_6 + x\,\cC_2\, F_4 -
(1{-}x)\,C_4\,\cF_2 +
\half(\mbox{$\frac{2}{3}$}{-}x)\,\im(\cC_2\,\bcF_2)\,\cF_2 \non \\
&&+\,\, \half(\third{-}x)\,\im(\cC_2\,\bcF_2)\,\cC_2\,. 
\eea
The Bianchi identities for these fields are
\bea
\label{wvbis}
&&\DD \cF_2 + i\,\bcF_2\,P + \cH_3 = 0\,, \non \\
&&\D F_4 + H_5 + \half\,\im(\bcF_2\,\cH_3) =0\,,\non \\
&&\DD \cF_6 + i\,\bcF_6 \,P + \cH_7 - x\,\cH_3\,F_4 +
(1{-}x)\,H_5\,\cF_2+\half(\mbox{$\frac{2}{3}$}{-}x)\,\cF_2\,\im(\cF_2\,\bcH_3)
= 0\,.
\eea
A crucial ingredient of supersymmetric brane actions is $\ka$-symmetry, a 
local world-volume symmetry for which the variation parameter $\ka$ is a 
target-space spinor satisfying $\ka=P_{+}\ze=\half(\id+\Ga)\ze$,
where $P_{+}$ is a projection operator of half-maximal rank.
It is generally accepted that the background theory being on shell is both a 
necessary and sufficient condition for $\ka$-invariance, 
although the necessity part has been explicitly proven only in a few cases. 
We will only investigate $\ka$-symmetry for on-shell backgrounds. 
The variations of the induced metric and the world-volume form fields under 
a $\ka$-transformation can be shown to be
\bea
\label{var}
\de_{\ka}g_{ij} &=&
2\,{E_{(i}}^{a}\,{E_{j)}}^{B}\,\ka^{\al}\,{T_{\al B}}^{b}\,\eta_{ab} +
{\rm c.c.}\,, \non \\
\de_{\ka}\cF_2 &=& -i\,\bcF_2\,i_{\ka}P-i_{\ka}\cH_3\,, \non \\ 
\de_{\ka}F_4 &=& -i_{\ka}H_5 + \half\,\im(\bcF_2\,i_{\ka}H_3)\,,\non \\
\de_{\ka}\cF_6 &=& -i\,\bcF_6\,i_{\ka}P - i_{\ka}\cH_7 +
x\,F_4\,i_{\ka}\cH_3 - (1{-}x)\,\cF_2\,i_{\ka}H_5 \non\\ && + \,\,
\half(\mbox{$\frac{2}{3}$}{-}x)\,\cF_2\,\im(\bcF_2\,i_{\ka}\cH_3)\,.
\eea 
The next step is to compute the variation of the action under a
$\ka$-transformation. On general grounds the action is taken
to be of the form
\be
\label{Ansatz}
S = \int \D^6 \xi\, \sqrt{-g}\,\la \left[ 1 + \Phi(\cF_2,\bcF_2,F_4)
- {*}\cF_6\,{*}\bcF_6\right]\,,
\ee
where $\la$ is a Lagrange multiplier for the constraint 
$\Ups = 1 + \Phi(\cF_2,\bcF_2,F_4) - {*}\cF_6\,{*}\bcF_6 \approx 0$.
For more details on actions of this type, see
refs~\cite{Bergshoeff:1992,Bergshoeff:1998c,Townsend:1997,Cederwall:1997,
Cederwall:1998a,Cederwall:1998b,Westerberg:1999}.
The function $\Phi$ is required to have U(1) charge zero but is
otherwise unconstrained at this stage.

It is often convenient to rewrite the action in ``form language'' as
\be
S = \int \la \left[ {*}1 + {*}\Phi(\cF_2,\bcF_2,F_4) +
\cF_6\,{*}\bcF_6\right]\,;
\ee
this form of the action is better suited for the derivation of the duality
relations supplementing it. 
These relations are constrained by compatibility 
with the equations of motion encoded in~(\ref{Ansatz}) and the Bianchi 
identities~(\ref{wvbis}) to take the form~\cite{Cederwall:1998a} 
\bea
\label{dualrel}
 -2x\,\re({*}\cF_{6}\,{*}\bcF_{2}) &=& K_{4} := \frac{\de\Phi}{\de
 F_4}\,,\non\\ (1{-}x)\,{*}\cF_{6}\,{*}F_{4} +
 \mbox{$\frac{i}{6}$}\,{*}[\re({*}\cF_6\,\bcF_2)\we \cF_2] &=&
 \cK_{2} := \frac{\de\Phi}{\de \bcF_2}\,
\eea
(for further details, see 
refs~\cite{Cederwall:1998a,Cederwall:1998b,Westerberg:1999}). 
These relations are a crucial ingredient in the 
$\ka$-symmetry analysis to be discussed next. As we
shall see, their complicated structure makes
this analysis difficult.

\setcounter{equation}{0}
\section{The method and the result}
\label{results}

In this section we determine the action and its associated duality
relations from the requirement of
$\ka$-symmetry. The analysis is significantly more complicated than for the 
cases considered previously in the literature as a consequence of the 
non-canonical structure of 
the duality relations (\ref{dualrel}) and the fact that the 
tension form is complex.
To make the problem tractable we 
will use a perturbative approach and expand the action in powers of the field 
strengths. At first sight it appears that adopting such a procedure would not 
be possible since there are identities which follow from the duality 
relations that mix terms of different orders. 
However, once these identities too are treated in a
perturbative order-by-order fashion the procedure becomes consistent. 

In order to establish the $\ka$-invariance of the action~(\ref{Ansatz}), 
it is sufficient to show that the variation of the constraint
$\Ups = 1 + \Phi(\cF_{2},\bcF_2,F_{4}) - {*}\cF_{6}\,{*}\bcF_6\approx0$ 
vanishes. Using a scaling argument, this variation is found to be
\bea
\de_{\ka}\Ups &=& 
(\bcK_2\scal\de_{\ka}\cF_2 + \cK_2\scal\de_{\ka}\bcF_2) 
+ K_4\scal\de_{\ka}F_4 + (\cF_6\scal\de_{\ka}\bcF_6+
\bcF_6\scal\de_{\ka}\cF_6) \non \\ && \hspace{-8mm}-
\bigg[\half\,({\bcK^{(i}}_{l}\,\cF^{j)l} + {\cK^{(i}}_{l}\,\bcF^{j)l})
 + \mbox{$\frac{2}{4!}$}\,{K^{(i}}_{lmn}\,F^{j)lmn} +
\mbox{$\frac{2\cdot3}{6!}$}\,\mbox{$\bcF^{(i}$}_{lmnpq}\,\cF^{j)lmnpq}
\bigg]\de_{\ka}g_{ij}\,. 
\eea
By inserting the explicit expressions (\ref{dualrel}) for the $K$'s, 
as well as the supergravity on-shell constraints given in eqs~(\ref{dim0})
and (\ref{dim1/2}), we obtain 
\bea
\label{variations}
(\de_{\ka}\Ups)^{(1/2)} &=& 
\bar{P}\Big[\bigg\{ i\,{*}\cF_6\,{*}\ga_6 - {*}[{*}\cF_6\,F_4
 +\mbox{$\frac{i}{2}$}\,\re({*}\cF_6 \bcF_2)\we \cF_2]\scal\ga_2\bigg\}\ka 
\non \\ 
&& \hspace{-1.1cm} + \,\bigg\{ i\,{*}\cF_6\,{*}[\cF_6 -(1{-}x)\,\cF_2\we F_4]
 + \sixth\,{*}[\re({*}\cF_6\,\bcF_2)\we\cF_2\we\cF_2] \bigg\} \bar{\ka}\Big]
 + {\rm c.c.}\,, \non\\
(\de_{\ka}\Ups)^{(0)} &=& 2i\,\bE_i \Big[\bigg\{ {*}\cF_6({*}\ga_5)^i 
-i{*}[{*}\cF_6\,F_4+\mbox{$\frac{i}{2}$}\,\re({*}\cF_6\,\bcF_2)\we\cF_2]^{ij}
\,\ga_j\bigg\}\bar{\ka} \non \\ 
&& \hspace{-1.1cm} + \, \bigg\{
{-}\mbox{$\frac{i}{6}$}\,\re({*}\cF_6\,{*}\bcF^{ijkl})\,\ga_{jkl}
+ \big[\{|{*}\cF_6|^2 + x\,{*}\re({*}\cF_6\,\bcF_2)\we F_4\}\,g^{ij} \non\\
&& \hspace{-1.1cm}
- \,\mbox{$\re({*}\cF_6\,\bcF_2)^{(i}$}_l\,{*}{F_4}^{j)l}\big]\ga_j
-\mbox{$\frac{i}{6}$}\,\im\big[{{*}\{\im({*}\cF_6\,\bcF_2)\we\cF_2\}^{(i}}_l\,
\bcF^{j)l}\big]\ga_j \bigg\}\ka\Big]+ {\rm c.c.}\,.
\eea
The next step is to insert the projected spinor parameter $\ka=P_+\ze$ into 
these variations using an appropriate Ansatz for the projection operator, 
and then examine the irreducible components of the expression obtained by 
expanding the products of $\ga$-matrices (for more details on the 
method and similar calculations see ref.~\cite{Westerberg:1999}). 
It turns out that the parameter $x$ is fixed to the value $\frac{2}{3}$ in 
the process, a value which corresponds to the field strengths
used in ref.~\cite{Cederwall:1998a} after taking into account some
differences
in conventions.
(Actually, it is difficult to conclusively rule out the possibility that $x$ 
could remain a free parameter; this would, however, require a very  
intricate form of $P_+$.) 
 
A major complication of the analysis arises from the fact that, in contrast
to all previously considered cases,  
an overall factor of the tension form can
not be factored out from the $\ka$-variation of the constraint. 
The reason for this can be traced to the fact that there are two
linearly independent tension forms (${*}\cF_6$ and ${*}\bcF_6$). 
This furthermore turns out to lead to the result that one does not get the 
duality relations in a simple form from any component; rather, one finds
the duality relations entangled with various identities implied by
 them. Although this makes the problem difficult,
it is still amenable to a perturbative approach, by means of which we have
determined the action and the associated projection operator to fourth
order in the world-volume fields. Higher-order corrections to the
action, if present, are expected to appear at sixth order only. 

The projection operator is found to be
\be
\label{projop}
2\,{*}\ga_6\,P_\pm\,\ze = {*}\ga_6\,\ze \mp
\bigg[\mbox{$\frac{2i}{3}$}\,{*}F_4\scal \ga_2\,\ze + \mbox{$\frac{i}{3}$}
\,{*}\cF_2\scal \ga_4\,\bar{\ze} + {*}\cF_6\,\bar{\ze} \bigg] + \cO(F^5)
\,,
\ee
with $\cO(F^5)$ denoting  terms of total
order five in the world-volume field strengths
 $\cF_2$, $\bcF_2$ and $F_4$. The final expression for the action is
\bea
\label{action}
S &=& \int\D^6\xi\,\sqrt{-g}\,\la\,\bigg[ 1 + \third \,\cF_2\scal \bcF_2 +
\mbox{$\frac{2}{3}$}\,F_4\scal F_4 +
\sixth(\beta{-}2)\,{*}(\cF_2\we F_4)\,{*}(\bcF_2\we F_4)
\non \\ &&\hspace{6mm}+\,\,
\sixth(1{-}\beta)\,(\cF_2\we\bcF_2)\scal({*}F_4\we{*}F_4) 
+ \sixth(\beta{-}\mbox{$\frac{2}{3}$})\,\cF_2\scal
\bcF_2 \, F_4\scal F_4 \non \\ &&\hspace{6mm}+\,\,
\sixth\,\beta\,(\cF_2\we{*}F_4)\scal (\bcF_2\wedge 
{*}F_4) + \cO(F^6) - {*}\cF_6\,{*}\bcF_6 \bigg]\,,
\eea
which is to be supplemented by the duality relations
\bea
\label{dualrels}
&&-\re({*}\cF_6\,{*}\bcF_2) = F_4 + \fourth(\beta{-}2)\,\re[{*}(\cF_2\we
F_4)\,{*}\bcF_2] + \fourth(1{-}\beta)\,{*}(\cF_2\we \bcF_2)\we{*}F_4\non\\
&&\qquad\qquad +\, \fourth(\beta{-}\mbox{$\frac{2}{3}$})\,
(\cF_2\scal \bcF_2) \, F_4 + \fourth \beta\,
\re[{*}(\cF_2\we{*}F_4)\we\bcF_2] + \cO(F^5) \,, \non\\
&&{*}\cF_6\,{*}F_4 + \mbox{$\frac{i}{2}$}\,{*}
[\re({*}\cF_6\,\bcF_2)\we\cF_2] = \cF_2
+ \half(\beta{-}2)\,(\cF_2\scal{*}F_4)\,{*}F_4 +
\half(\beta{-}\mbox{$\frac{2}{3}$})\,(F_4\scal F_4) \, \cF_2  \non \\ && 
\qquad\qquad -\, \half(1{-}\beta)\,{*}[{*}({*}F_4\we{*}F_4)\we \cF_2]-
\half\beta\,{*}[{*}({*}F_4\we\cF_2)\we{*}F_4] + \cO(F^5)\,. 
\eea
Here $\beta$ is a free parameter (see below).  
Two ubiquitous identities in the $\ka$-symmetry calculations are
$[\cF_2,\bcF_2]=0$ and $[\cF_2,{*}F_4]=0$, where $\cF_2$, $\bcF_2$ and
${*}F_4$ are viewed as matrices and the bracket is a matrix
commutator. Another important result needed to verify $\ka$-symmetry is
the relation
\be
\im\,({*}\cF_6\,{*}\bcF_2) = -\sixth\cF_2\we\bcF_2 +
\mbox{$\frac{2}{3}$}\,{*}F_4\we {*}F_4 + \cO(F^6)\,.
\ee
These identities can be shown to follow from the duality relations
(\ref{dualrels}). They are also required in order for $P_+$ to have
the correct properties. In addition, one  needs to use the
fact that the following relation holds when the 
duality relations are satisfied: 
\be
\label{phid}
\Phi \approx \mbox{$\frac{1}{9}$}\,\cF_2\scal\bcF_2 +
\mbox{$\frac{4}{9}$}\,F_4\scal F_4 + \cO(F^6)\,.
\ee
This relation follows from the expression for $\Phi$ encoded in
(\ref{action}) combined with 
the identity $2x\,\re \,(\cK_2\scal\bcF_2) +
(1{-}x)\,K_4\scal F_4 = 0$ (for $x=\frac{2}{3}$), which
 can readily be derived from the form of the duality
relations (\ref{dualrel}). Once one has shown that the duality
relations imply the above commutator identities, it is straightforward
to check that the terms in the duality relations proportional to
$\beta$ vanish for purely algebraic reasons, 
showing that $\beta$ can chosen arbitrarily. (It is likely that $\beta$ will be
fixed in the complete action.) Let us also mention that one can change
the appearance of the fourth-order
terms. For instance, it follows from the duality relations given above
that $\cF_2 \we \bcF_2 -{*}F_4\we{*}F_4 = \cO(F^4)$; adding this
expression squared to the action does not violate $\ka$-symmetry to
fourth order.  

In order to see that the  tension of the $(p,q)$ branes described by the 
 action (\ref{action}) works out correctly one proceeds analogously to
the discussion in 
ref.~\cite{Cederwall:1997}.  
To get agreement with the formula for the tension first 
obtained in ref.~\cite{Witten:1995b}, one has to take into account the
 fact that
 when transforming to the string frame the tension receives an
 additional 
 overall factor 
$e^{-\phi}$ compared to the string case (recall that $g^{\rm string}_{mn} =
 e^{\frac{1}{2}\phi}g^{\rm Einstein}_{mn}$).

The question arises to what extent the above
action differs from the
complete action. It appears likely that the modifications, if any, should be 
rather minor. Furthermore, it is not clear whether
$P_+$ has to be modified (the above expression, obtained from a fairly
general Ansatz, certainly looks
deceptively simple). However, if (\ref{projop}) is the complete
result, it seems
difficult to modify the action without ruining the property~(\ref{phid}). 
It appears that some new input is needed to make further progress; this is
especially true in the case of the search for a manifestly SL(2,$\Z$)-covariant
formulation of the seven-branes. These are known~\cite{Meessen:1998} to 
form a triplet and couple to the  eight-form potentials dual to the
three scalars which belong to the SL(2,$\R$)/U(1) coset. Perhaps one way 
to make further progress is via T-duality; it may be possible to derive
T-duality rules which relate duality covariant actions in the 
M/type {\II}A and type {\II}B theories, and in this way make the problem more
tractable.
It would also be desirable to have a more uniform description of the 
manifestly SL(2,$\Z$)-covariant type-{\II}B-brane actions.

\subsection*{Acknowledgements}

The work of A.W. and N.W. was supported by the European Commission under
the contracts FMBICT972021 and FMBICT983302, respectively.
A.W. would like to thank Martin Cederwall for collaboration
on the project which initiated the present study.

\appendix
\setcounter{equation}{0}
\section{Conventions}
\label{appA}

We employ a complex superspace notation in which a one-form is expanded in
a local inertial-frame basis as 
$\Om_1=E^A\,\Om_A=E^a\,\Om_a+E^\al\,\Om_\al+E^{\bar{\al}}\,\Om_{\bar{\al}}$,
 with $E^{\bar{\al}} = \overline{E^{\al}}$. The relation to the 
real formulation used in 
ref.~\cite{Cederwall:1996c} is  $E^{\al} = E^{1\al}{+}
iE^{2\al}$, $E^{\bar{\al}} = E^{1\al} {-} iE^{2\al}$,  
$\Om_{\al} = \half(\Om_{1\al} {-}
i\,\Om_{2\al})$ and $\Om_{\bar{\al}} = \half(\Om_{1\al} {+}
i\,\Om_{2\al})$. Given these translation rules our spinor conventions
follow those of ref. \cite{Cederwall:1996c}. Moreover, complex
conjugation 
of a bispinor reverses the order of the spinors.  
For higher forms we use the additional convention
$\Om_{n}  
=\mbox{$\frac{1}{n!}$}E^{A_n}\we\ldots\we E^{A_1} \Om_{A_1\ldots
A_n}$. 
The exterior derivative $\D$ acts from the right, so that
$\D(\Om_m\we\tilde{\Om}_n) = \Om_m\we\D\tilde{\Om}_n 
+(-1)^n\D\Om_m\we\tilde{\Om}_n$.
(We usually suppress the symbol $\we$ when no confusion should arise.)
The world-volume forms (which are bosonic) follow the same conventions and 
hence obey the same rules. Furthermore, we do not distinguish notationally 
between a target-space form $\Om_n$ and its pull-back to the world-volume, 
the components of which are given by
\be
\Om_{i_1\ldots i_n}=
{E_{i_n}}^{A_n}\ldots{E_{i_1}}^{A_1}\,\Om_{A_1\ldots A_n} :=
\pa_{i_n}Z^{M_n}\,{E_{M_n}}^{A_n}\ldots\pa_{i_1}Z^{M_1}\,{E_{M_1}}^{A_1}\,
\Om_{A_1\ldots A_n}\,.
\ee
The Hodge dual of a world-volume $n$-form is defined by
\be
({*}\Om_n)^{i_{1}\ldots i_{6-n}} = \mbox{$\frac{1}{n!\sqrt{-g}}$}
\ep^{i_1\ldots i_6} \Om_{i_{6-n+1}\ldots i_{6}}\,,
\ee
where $g$ is the determinant of the induced metric $g_{ij}{=}{\pa_i
Z}^m{\pa_j Z}^n \,g_{mn}$ (with mostly-plus signature) 
and $\ep^{i_1\ldots i_6}$ is the 
totally antisymmetric tensor density satisfying $\ep^{01\ldots5}=+1$. 
World-volume $\ga$-matrices are defined as the pull-backs 
$\ga_i={E_i}^a\,\Ga_a$. Their symmetrised product obeys the Clifford algebra 
$\{\ga_i,\ga_j\}=2\,g_{ij}\id$ inherited from the target-space, while their
antisymmetrised products can be combined into the forms
\be
\ga_n=\mbox{$\frac{1}{n!}$}\,\D\xi^{i_n}\we\ldots\we\D\xi^{i_1}\,
\ga_{i_1\ldots i_n}\,,
\ee
where $\ga_{i_1\ldots i_n}=\ga_{[i_1}\ldots\ga_{i_n]}$ and the 
antisymmetrisation is of weight one. And finally, we use the notation
\be
A_n\scal B_n = \mbox{$\frac{1}{n!}$}A^{i_1\ldots i_n}B_{i_1\ldots i_n}
\ee
for the scalar product of two world-volume $n$-forms.

\begingroup\raggedright\endgroup


\begin{thebibliography}{10}

\bibitem{Hull:1995}
C.~M. Hull and P.~K. Townsend, ``Enhanced gauge symmetries in superstring
  theories.'' Nucl. Phys. {\bf B451} (1995) 525,
  {{\tt hep-th/9505073}}.

\bibitem{Schwarz:1995}
J.~H. Schwarz, ``An SL(2,$\Z$) multiplet of type IIB superstrings.'' Phys.
  Lett. {\bf B360} (1995) 13; erratum ibid. {\bf B364} (1995) 252,
  {{\tt hep-th/9508143}}.

\bibitem{Witten:1995b}
E.~Witten, ``Bound states of strings and $p$-branes.'' Nucl. Phys. {\bf B460}
  (1996) 335, {{\tt hep-th/9510135}}.

\bibitem{Duff:1991}
M.~J. Duff and J.~X. Lu, ``The selfdual type IIB superthreebrane.'' Phys. Lett.
  {\bf B273} (1991) 409.

\bibitem{Lu:1998}
J.~X. Lu and S.~Roy, ``An SL(2, $\Z$) multiplet of type IIB super
  five-branes.'' Phys. Lett. {\bf B428} (1998) 289,
  {{\tt hep-th/9802080}}.

\bibitem{Meessen:1998}
P.~Meessen and T.~Ort{\'\i}n, ``An SL(2,$\Z$) multiplet of nine-dimensional
  type II supergravity theories.'' Nucl. Phys. {\bf B541} (1999) 195,
  {{\tt hep-th/9806120}}.

\bibitem{Howe:1984}
P.~S. Howe and P.~C. West, ``The complete $N=2$, $d=10$ supergravity.'' Nucl.
  Phys. {\bf B238} (1984) 181.

\bibitem{Dall'Agata:1998}
G.~Dall'Agata, K.~Lechner, and M.~Tonin, ``$D = 10$, $N =$ IIB supergravity:
  Lorentz invariant actions and duality.'' JHEP {\bf 9807} (1998) 017,
  {{\tt hep-th/9806140}}.

\bibitem{Townsend:1997}
P.~K. Townsend, ``Membrane tension and manifest IIB S duality.'' Phys. Lett.
  {\bf B409} (1997) 131, {{\tt
  hep-th/9705160}}.

\bibitem{Cederwall:1997}
M.~Cederwall and P.~K. Townsend, ``The manifestly Sl(2;$\Z$) covariant
  superstring.'' JHEP {\bf 9709} (1997) 003,
  {{\tt hep-th/9709002}}.

\bibitem{Cederwall:1998a}
M.~Cederwall and A.~Westerberg, ``World volume fields, SL(2;$\Z$) and duality:
  the type IIB three-brane.'' JHEP {\bf 9802} (1998) 004,
  {{\tt hep-th/9710007}}.

\bibitem{Cederwall:1998b}
M.~Cederwall, B.~E.~W. Nilsson, and P.~Sundell, ``An action for the
  superfive-brane in $D{=}11$ supergravity.'' JHEP {\bf 9804} (1998) 007,
  {{\tt hep-th/9712059}}.

\bibitem{Westerberg:1999}
A.~Westerberg and N.~Wyllard, ``Supersymmetric brane actions from interpolating
  dualisations.'' {{\tt
  hep-th/9904117}}.

\bibitem{Schwarz:1983}
J.~H. Schwarz, ``Covariant field equations of chiral $N=2$ $D=10$
  supergravity.'' Nucl. Phys. {\bf B226} (1983) 269.

\bibitem{Cederwall:1996c}
M.~Cederwall, A.~von Gussich, B.~E.~W. Nilsson, P.~Sundell, and A.~Westerberg,
  ``The Dirichlet super $p$-branes in ten-dimensional type IIA and IIB
  supergravity.'' Nucl. Phys. {\bf B490} (1997) 179,
  {{\tt hep-th/9611159}}.

\bibitem{Bergshoeff:1992}
E.~Bergshoeff, L.~A.~J. London, and P.~K. Townsend, ``Space-time scale
  invariance and the super $p$-brane.'' Class. Quant. Grav. {\bf 9} (1992)
  2545, {{\tt hep-th/9206026}}.

\bibitem{Bergshoeff:1998c}
E.~Bergshoeff and P.~K. Townsend, ``Super D-branes revisited.'' Nucl. Phys.
  {\bf B531} (1998) 226, {{\tt hep-th/9804011}}.

\end{thebibliography}
\end{document}